\documentclass[epj,referee]{svjour}
\usepackage{graphics}
\usepackage{amssymb,amsmath}
\begin{document}
\title{Damage spreading in 2-dimensional isotropic and anisotropic Bak-Sneppen models}
\author{Burhan Bakar\thanks{\emph{E-mail:} burhan.bakar@mail.ege.edu.tr} \and Ugur Tirnakli\thanks{\emph{E-mail:} ugur.tirnakli@ege.edu.tr} 
}                     
%
%
\institute{Department of Physics, Faculty of Science, Ege University, 35100 Izmir, Turkey}
\date{Received: date / Revised version: date}
%
\abstract{
We implement the damage spreading technique on $2$-dimensional isotropic and anisotropic Bak-Sneppen models. Our extensive numerical simulations show that there exists a power-law sensitivity to the initial conditions at the statistically stationary state (self-organized critical state). Corresponding growth exponent $\alpha$ for the Hamming distance and the dynamical exponent $z$ are calculated. These values allow us to observe a clear data collapse of the finite size scaling for both versions of the Bak-Sneppen model. Moreover, it is shown that the growth exponent of the distance in the isotropic and anisotropic Bak-Sneppen models is strongly affected by the choice of the transient time.     
\PACS{
      {05.65.+b}{Self-organized systems}   \and
      	{64.60.Ht}{Dynamic critical phenomena} \and
      	{87.23.Kg}{Dynamics of evolution}
     } 
} 
\maketitle
\section{\label{sec:Int}Introduction}
In 1993, Bak and Sneppen (BS) introduced a simple model to describe the biological evolution of an ecology of interacting species \cite{BSPRL-71}. Since then, it has attracted quite some attention among the natural scientists as well as economists. This wide range of interest comes from the fact that BS model is the simplest model that exhibits self-organized criticality \cite{BSPRL-71,BSWPRL59}. The self-organized criticality feature of the BS model is revealed in its ability to naturally evolve towards a scale invariance stationary state \cite{PRE53-414}. That is, the  correlation length in the BS model is infinite and an initial local perturbation might lead to a global effect. Therefore, it is worth to study the sensitivity to the initial conditions in the BS model.  

The technique that we use to study the sensitivity to the initial conditions in the BS model is known as the {\it damage spreading technique} in dynamical systems theory and can be described as follows:
If we consider two copies of the same dynamical system starting from slightly different initial conditions and follow their time evolution, the sensitivity function can be defined as,
\begin{equation}\label{eq:sens}
\xi(t)\equiv\lim_{\Delta x(0)\rightarrow0}\frac{\Delta x(t)}{\Delta x(0)}=\exp{(\lambda t)},
\end{equation}  
where $\lambda$ is the Lyapunov exponent and $\Delta x(0)$ and $\Delta x(t)$ are the distances between two copies at $t=0$ and $t$, respectively. Depending on the $\lambda$ being positive, negative or zero three different behavior can be distinguished from Eq.~(\ref{eq:sens}): (i) $\lambda>0$, the system is said to be strongly sensitive to the initial conditions, (ii) $\lambda<0$, the system is said to be strongly insensitive to the initial conditions, (iii) $\lambda=0$, while the sensitivity function could be a whole class of functions, for the low-dimensional discrete dynamical systems Eq.~(\ref{eq:sens}) forms a power-law, 
\begin{equation}
\xi(t)\sim t^{\alpha},
\end{equation}    
where $\alpha$ is some exponent and $\alpha>0$ and $\alpha<0$ cases correspond to weakly sensitive and weakly insensitive to the initial conditions, respectively \cite{alpha1,alpha4,alpha2,alpha5}. For the high dimensional systems like the BS model, the same analysis could be performed using the so-called {\it Hamming distance} instead of the sensitivity function. 

Our task will be to investigate the short-time dynamics of the isotropic and anisotropic BS models on a square lattice using the standard damage spreading technique. As far as we know, this technique has not been used in the literature to investigate the sensitivity to the initial conditions in  $2$-dimensional ($2d$) BS model while there have been several works in $1d$ \cite{TsallisEPJB1,EPJB4,JPA32,EPJB7,UTPhysica344}. In $2d$, with extensive simulations, it is our hope to obtain similar behavior of the temporal and spatial correlation functions with $1d$ isotropic and anisotropic BS models. Furthermore, we expect topologically the same short-time evolution of the Hamming distance with different scaling exponents. In order to identify the dynamical quantities in $2d$ BS model we will introduce in the second section the well known isotropic and anisotropic versions of the BS model and compare our results for the critical exponent values of temporal and spatial correlations to the known findings. In the third section, we implement the damage spreading technique on $2d$ isotropic and anisotropic BS models and calculate the related critical exponents. The relation between some model parameters and the critical exponents is reviewed. Moreover, we analyze the finite size scaling of the normalized Hamming distance for both versions of the BS model. Finally, in the last section, we discuss the results and give some perspectives.
\section{Isotropic and anisotropic versions of BS model}
The dynamics of BS model on a $2$-dimensional lattice of an edge size $L$ has simple rules. The initial state of the system on a square lattice is characterized by $N=L\times L$ fitness values (random numbers) $f_{i,\,j}$, where $i=1\cdots L$ and $j=1\cdots L$, uniformly distributed between $0$ and $1$. These fitness values are assigned to each site $i,\,j$ of the lattice with periodic boundary conditions.  The usual dynamics of the system is eventually achieved by localizing the lattice site with the minimum fitness $f_{min}$ and assigning new random numbers to that site and its first nearest neighbors. 

The model described above can be called as an {\it isotropic} BS model since the interaction between the current minimum and its first nearest neighbors is the same in both directions. In other words, if one observes the minimum fitness value of the system at time step $t$ as $f_{min}=f_{i,\,j}$ then the fitness values $f_{i-1,\,j}$, $f_{i,\,j-1}$, $f_{i,\,j}$, $f_{i+1,\,j}$ and $f_{i,\,j+1}$ will be updated at that time step. This means that the possibility for the minimum to jump to one of its left or right nearest neighbors or to one of its up or down nearest neighbors in the next time step is simply the same, with a resulting isotropic avalanche of events.

On the other hand, one can easily consider alternative updating rules. One possible alternative could be to update at each time step $f_{i-u,\,j}$, $f_{i,\,j-\ell}$, $f_{i,\,j}$, $f_{i+b,\,j}$ and $f_{i,\,j+r}$ (where $u$, $\ell$, $b$ and $r$ are arbitrary positive integers taken from the interval $[1,\,L]$). Let us consider, for example, the case of  $f_{i-2,\,j}$, $f_{i,\,j-1}$, $f_{i,\,j}$, $f_{i+1,\,j}$ and $f_{i,\,j+2}$. The system now has an inherent bias to the up and right. This means that we would expect a preferred direction for an avalanche to propagate. Therefore, this model is called as the {\it anisotropic} BS model. In principle, several types of anisotropy can be introduced  by changing the values of $u$, $\ell$, $b$ and $r$, provided that $\ell\neq r$ and $u\neq b$ (since $\ell=r$ and $u=b$ represents the isotropic BS model). The maximal anisotropic cases are defined by $\ell=0$ ($\ell=1$), $r=1$ ($r=0$) and $b=0$ ($b=1$), $u=1$ ($u=0$) whereas other definitions are considered as intermediate anisotropies. In this work, we use the maximal anisotropic case in all simulations since the convergence is faster than the intermediate anisotropy choices \cite{JPA31}. 
 
After some transient time which depends on the size of the system the isotropic and anisotropic models achieve a statistically stationary state (i.e., self-organized critical state) in which the density of fitness values is uniformly distributed on $[f_{c},\,1]$ and vanishes on $[0,\,f_{c}]$, where the critical threshold value $f_{c}$ depends on the lattice dimension. On the 2-dimensional lattice considered here, $f_{c}\simeq0.328$ \cite{PRE53-414} for the isotropic BS model and $f_{c}\simeq0.439$ for the anisotropic BS model. The difference between the critical threshold values of isotropic and anisotropic BS models comes from the change in the rates of the spreading out of avalanches. Once the stationary state is achieved, the temporal and spatial correlation functions are power-law in both models signifying the existence of a critical state with no characteristic length or time scales  (scale invariance). These correlations can be used to determine the universality classes of such models. The distribution of the absolute distance $x$ between successive minima is a good example for the spatial correlation and defined by $P_{jump}(x)\sim x^{-\pi}$, where $\pi\simeq2.92$ for the isotropic BS model and $\pi\simeq2.57$ for the anisotropic BS model on a square lattice. The temporal correlations $P_{first}(t)$, the distribution of first return times, and $P_{all}(t)$, the distributions of all return times scale as $P_{first}(t)\sim t^{-\tau_{first}}$ and $P_{all}(t)\sim t^{-\tau_{all}}$, where $\tau_{first}\simeq1.24$ and $\tau_{all}\simeq0.71$ for the isotropic BS model; $\tau_{first}\simeq1.32$ and $\tau_{all}\simeq0.85$ for the anisotropic BS model. These different values of $\pi$, $\tau_{first}$ and $\tau_{all}$ suggest that the isotropic and anisotropic BS models belong to different universality classes. Our simulation results for the critical exponent values of temporal and spatial correlations given above are in good agreement with the known results for $2d$ BS model \cite{PRE53-414,JPA31,PRL80,PRE58}.
\begin{figure*}\sidecaption
\resizebox{\hsize}{!}{\includegraphics*{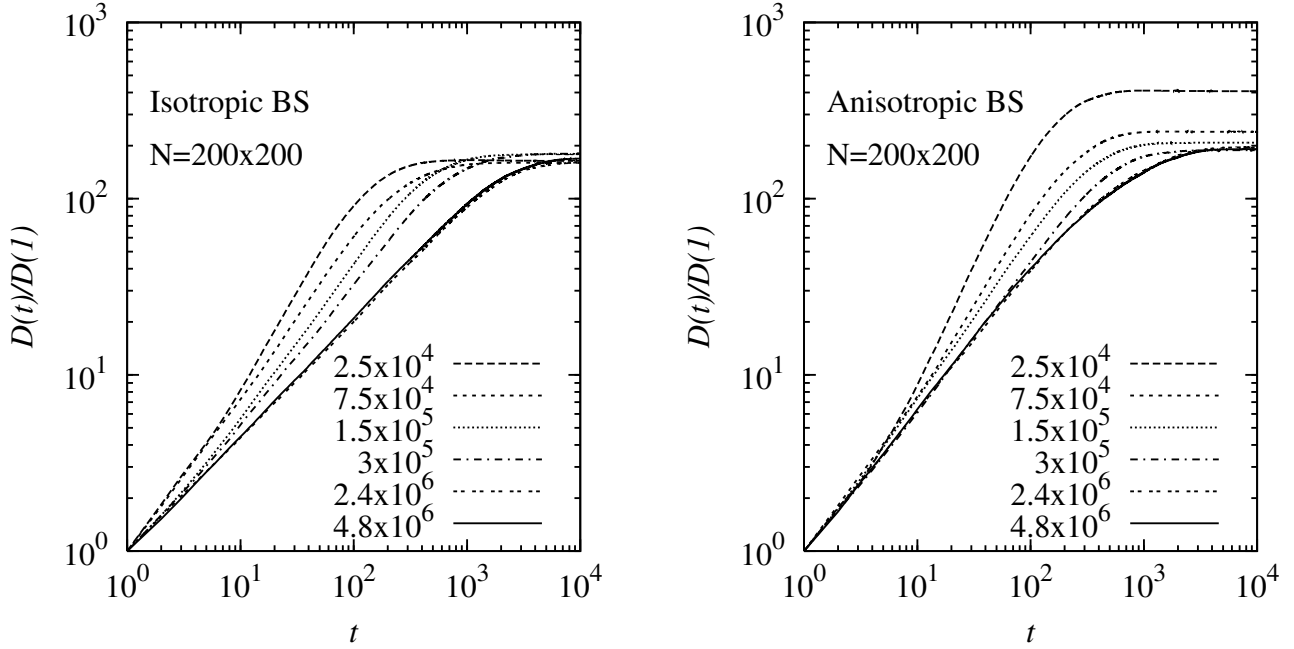}}
\caption{Time evolution of the normalized Hamming distance for different transient times of the isotropic (left) and anisotropic (right) BS models on a square lattice with $N=200\times200$ fitness values. As the transient time increases, the slope value decreases and eventually, converges on a fixed value. For  $N=200\time200$ the slope values are $\alpha\simeq0.66$ and $\alpha\simeq0.74$ for the isotropic and anisotropic BS models, respectively.}
\label{fig:transients}
\end{figure*}
\section{\label{sec:DS}Damage spreading}
As mentioned in the Sec. (\ref{sec:Int}), the sensitivity to the initial conditions of the isotropic and anisotropic BS models on a square lattice can be studied by damage spreading technique. This technique has been used in the literature previously to investigate the propagation of local perturbations in $1d$ BS model \cite{TsallisEPJB1,EPJB4,JPA32,EPJB7,UTIJMPC14,UTPhysica342,UTPhysica344}. In these works, measuring the evolution of the discrepancy between two initially close configurations under the same noise, it was shown that this distance exhibits an initial power-law divergence, followed by a finite size-dependent saturation regime.

The algorithm for the BS model on a square lattice with $N=L\times L$ fitness values can be introduced as follows: 
\begin{enumerate}
\item Once the stationary state has been achieved, consider the system as replica $1$ denoted by $f_{i,\,j}^{1}$.
\item Produce an identical copy of $f_{i,\,j}^{1}$ and introduce a small damage in this copy by interchanging the site with minimum fitness with a randomly chosen site (denote the new replica as $f_{i,\,j}^{2}$).
\item Let both replicas ($f_{i,\,j}^{1}$ and $f_{i,\,j}^{2}$) evolve in time using always the same set of random numbers. 
\end{enumerate}
\begin{figure*}\sidecaption
\resizebox{\hsize}{!}{\includegraphics*{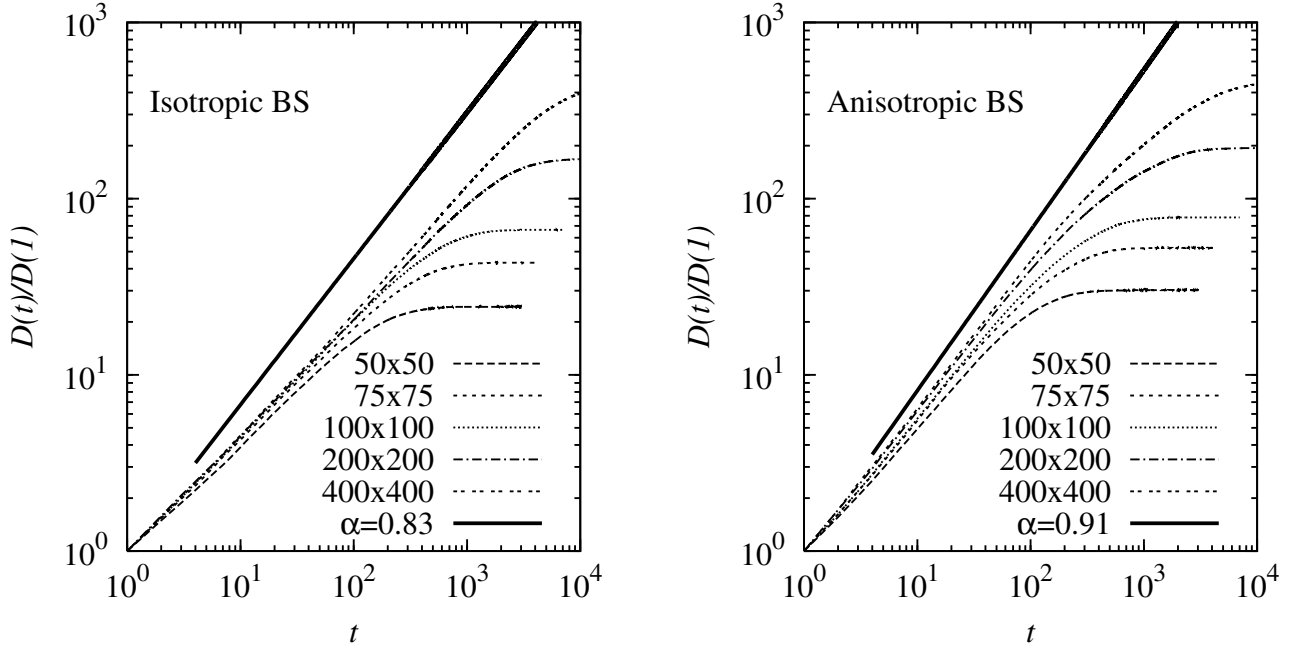}}
\caption{Time evolution of the normalized Hamming distance for five different system sizes of the isotropic (left) and anisotropic (right) BS models. In $N\rightarrow\infty$ limit, the slope values are estimated as $\alpha=0.83\pm0.03$ for the isotropic (left) and $\alpha=0.91\pm0.03$ for the anisotropic case (right), respectively (see Fig.~(\ref{fig:ConfInt})).}
\label{fig:ShortD0}
\end{figure*}
We can now define the Hamming distance between two replicas as,
\begin{equation}\label{eq:Hamming}
D(t)=\left\langle\frac{1}{N}\sum_{i,\,j=1}^{L}\lvert f_{i,\,j}^{1}-f_{i,\,j}^{2}\rvert\right\rangle,
\end{equation}
where $\langle\cdots\rangle$ stays for the configurational averages over various realizations.    Although, one of us introduced a new definition of the Hamming distance for the damage spreading technique \cite{UTIJMPC14,UTPhysica342,UTPhysica375}, which allows one to analyze both short- and long-time dynamics, in this work we use the standard definition given in (\ref{eq:Hamming}), since we aim to analyze the short-time dynamics better by performing extensive simulations.     
\begin{figure*}\sidecaption
\resizebox{0.90\hsize}{!}{\includegraphics*{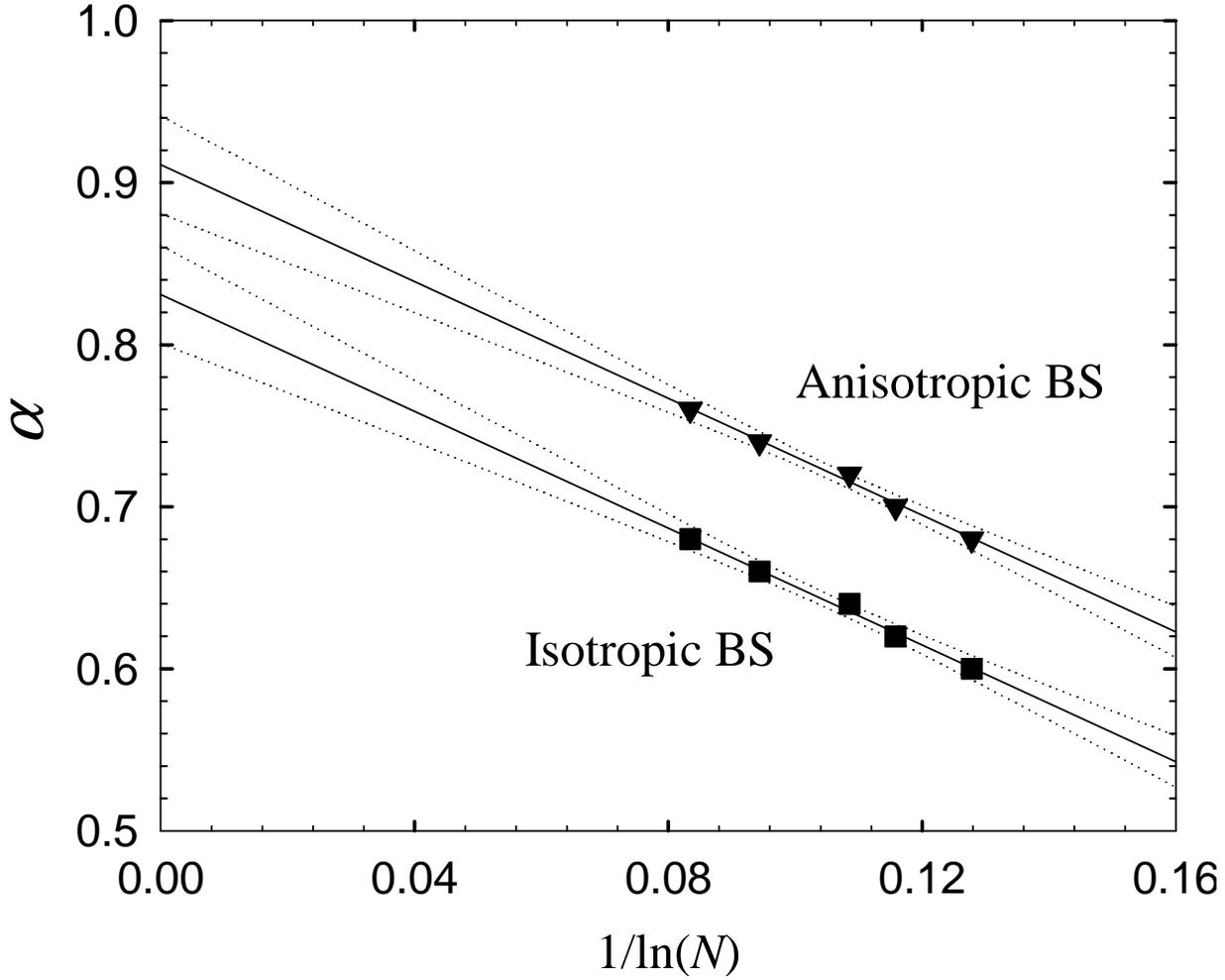}}
\caption{The confidence interval of $\alpha$ in $N\rightarrow\infty$ limit. From this scaling we estimate the growth exponent of Hamming distance as $\alpha=0.83\pm0.03$ and $\alpha=0.91\pm0.03$ for the isotropic ($\blacksquare$) and anisotropic ($\blacktriangledown$) BS  models, respectively.}
\label{fig:ConfInt}
\end{figure*}

In our simulations, all the results are obtained by averaging over $800$ realizations, which is enough to avoid the 
fluctuations. The largest system size that we achieve is $N=400\times400=1.6\times10^{5}$ for both isotropic and 
anisotropic versions of the BS model. In order to let the system achieve the statistically stationary state we choose 
an appropriate transient time for each lattice size. The choice of the number of transients is important to obtain the 
most realistic slope value. For a fixed lattice size, one can attain a couple of different transient times each of 
which is enough to observe the power-law growth of the Hamming distance. But, as it can be seen 
from Fig.~(\ref{fig:transients}), the smallest transient time gives the largest exponent. 
As the transient time increases, the exponent value decreases until it reaches a fixed value and further increment of 
the transient time does not affect the slope anymore. This is the most reliable way of choosing the appropriate 
transient time for each system size. Otherwise one can easily overestimate the exponent of the power-law growth. 
In our simulations, we choose this number as $4\times10^{5}$, $7\times10^{5}$, $1\times10^{6}$, $2.4\times10^{6}$ and 
$6.4\times10^{6}$ for $50\times50$, $75\times75$, $100\times100$, $200\times200$ and $400\times400$ lattices, 
respectively. Once the stationary state is achieved by attaining the most appropriate transient time, one can study 
the short-time dynamics of the model using the measure given in Eq.~(\ref{eq:Hamming}). For $1d$ isotropic and 
anisotropic BS models, it is known from the literature that this measure exhibits initially a power-law growth such 
as $D_{0}(t)\sim t^{\alpha}$ and saturates at a constant 
value \cite{TsallisEPJB1,EPJB4,JPA32,EPJB7,UTIJMPC14,UTPhysica342,UTPhysica344}. As it is evident from 
Fig.~(\ref{fig:ShortD0}), our simulation results claim that the procedure that describes the time evolution of the 
Hamming distance in $1d$ isotropic and anisotropic BS models remains true also for $2d$ isotropic and anisotropic 
BS models. The limiting case exponents are obtained as $\alpha\simeq0.83\pm0.03$ and $\alpha\simeq0.91\pm0.03$ for 
the isotropic and anisotropic versions of the BS model, respectively using the extrapolation in Fig.~(\ref{fig:ConfInt}). These results of $2d$ can be compared to the ones coming from the $1d$ cases (namely, $\alpha=0.48$ and $\alpha=0.53$ 
for the $1d$ isotropic and $1d$ anisotropic cases, respectively \cite{UTPhysica344}). The increase of the exponent as the 
dimension increases, for both isotropic and anisotropic cases, reveals that the ability of the system to cover the whole 
lattice increases with the dimension. It is also possible to compare these results with the mean-field values ($\sim1$) 
obtained in \cite{JPA32} for the ring model and in \cite{UTEPJ46} for the coherent noise model.   
\begin{figure*}\sidecaption
\resizebox{0.90\hsize}{!}{\includegraphics*{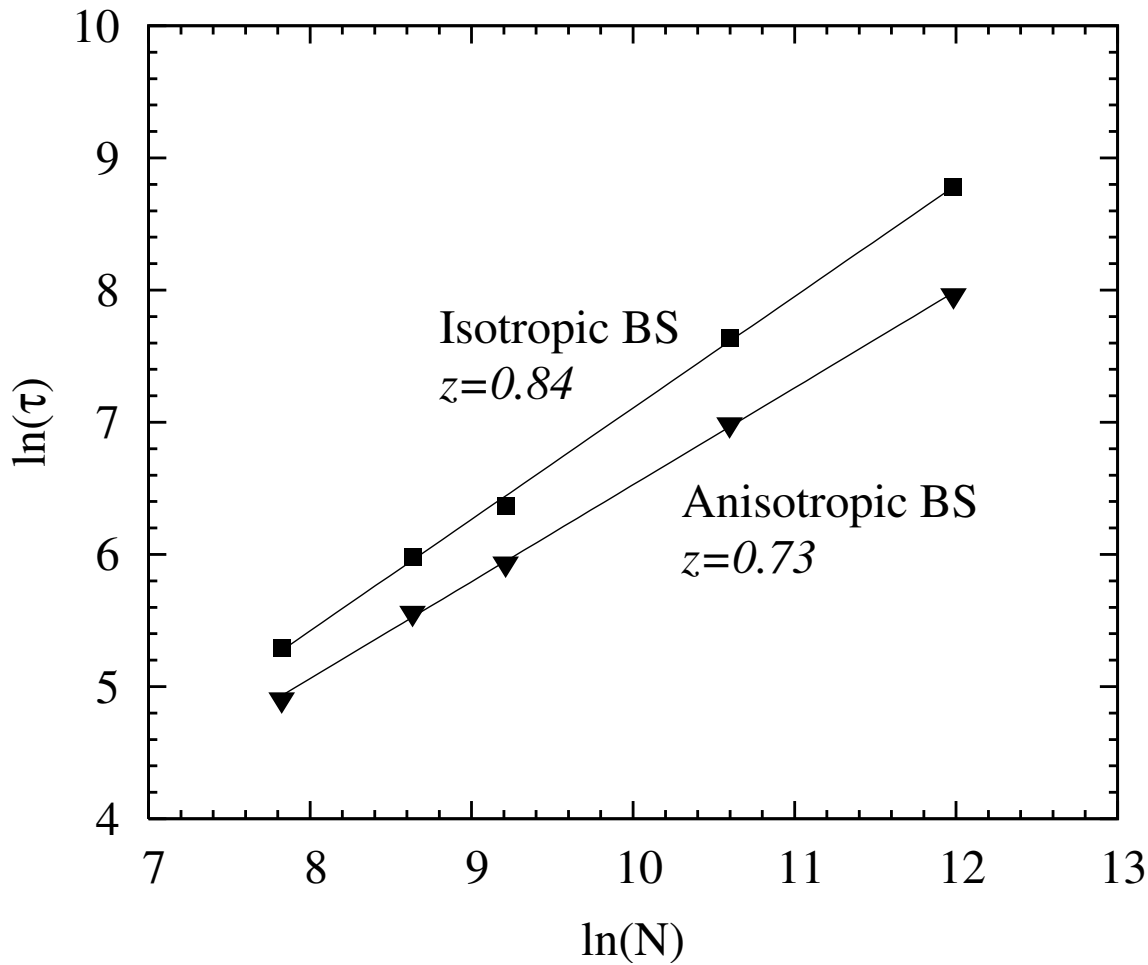}}
\caption{Log-log plot of $\tau$ versus $N$ for the five curves given in Fig.~(\ref{fig:ShortD0}). The estimated dynamical exponent values are $z=0.84$ and $z=0.73$ for the isotropic ($\blacksquare$) and anisotropic ($\blacktriangledown$) BS models, respectively.}
\label{fig:DynExp}
\end{figure*}
\subsection{Dynamical exponent and finite size scaling}    
The second important exponent defined by $z$ is called the dynamical exponent and comes from the scaling of $\tau(N)\sim N^{z}$, where $\tau$ is defined to be the value of $t$ at which the power-law increasing part of the Hamming distance measurement crosses over onto the saturation regime for fixed $N$ (namely, in Fig.~(\ref{fig:ShortD0}), intersection of two straight lines drawn through the linearly increasing power-law curve and the horizontal constant plateau).  From Fig.~(\ref{fig:DynExp}), we obtained the dynamical exponent as $z\simeq0.84$ and $z\simeq0.73$ for the isotropic and anisotropic versions, respectively. 
\begin{figure*}\sidecaption
\resizebox{\hsize}{!}{\includegraphics*{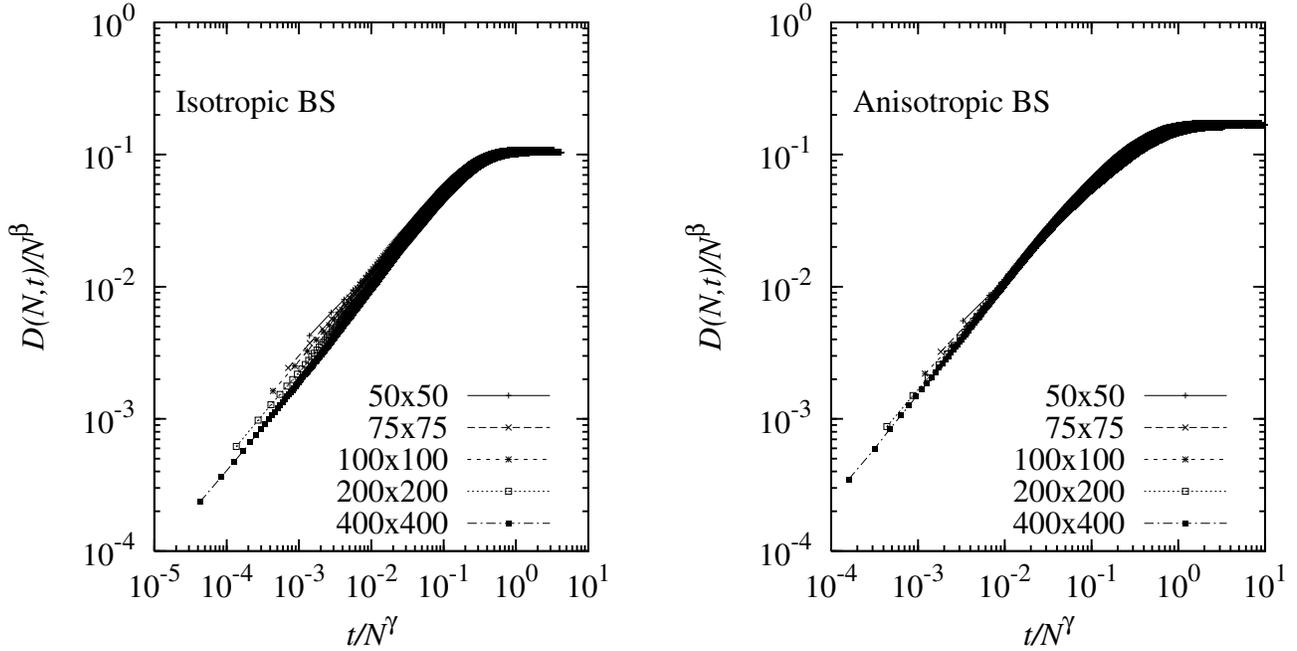}}
\caption{Data collapse of finite size scaling given in Eq.~(\ref{eq:FiniteSc}) for the five curves of Fig.~(\ref{fig:ShortD0}) in the isotropic (left) and anisotropic (right) BS models.}
\label{fig:DataColl}
\end{figure*}

The obtained exponents $\alpha$ and $z$ can be used to analyze the finite size scaling behavior of the isotropic and anisotropic BS models on a square lattice. Namely, using the exponent values one can control if the normalized Hamming distance $D(N,t)=\langle D(t)/D(1)\rangle$ obeys the following finite size scaling behavior,
\begin{equation}\label{eq:FiniteSc}
D(N,t)\sim N^{\beta}F\left(\frac{t}{N^{\gamma}}\right),
\end{equation}   
where $\beta=\alpha\gamma$ and $\gamma=z$ ($\beta\simeq0.69$, $\gamma\simeq0.84$ and $\beta\simeq0.66$, $\gamma\simeq0.73$ for the isotropic and anisotropic cases, respectively). Substitution of corresponding exponent values in Eq.~(\ref{eq:FiniteSc}) leads to a clear data collapse for each version of BS model, as it can be seen in Fig.~(\ref{fig:DataColl}). Different values of the dynamical exponent obtained for isotropic and anisotropic cases as well as $1d$ and $2d$ cases indicate that the dynamical property, characterized by the time needed to cover all lattice sites, differs for each case. These values are also different from the mean-field model results ($\sim1$) \cite{JPA32,UTEPJ46}, which clearly states that the time for the system to reach the plateau scales linearly with the lattice size.


\section{Discussion and conclusion}
We study the short-time dynamics of $2d$ isotropic and anisotropic BS models on a square lattice. Our extensive simulation results for the power-law growth of the temporal and spatial correlations are in good agreement with the known results of the previous works. Moreover, the time evolution of the Hamming distance is studied by implementing the damage spreading technique and it is revealed that $2d$ isotropic and anisotropic BS models exhibit weak sensitivity to the initial conditions as it is the case for $1d$ BS model on a chain. It is shown that the scaling of the Hamming distance strongly depends on the choice of the transient time. That is, if one chooses a relatively large transient time for a fixed system size $N$ to have the system at the statistically stationary state, a power-law behavior of the Hamming distance in time can be observed. But it is seen that increasing the transient time leads to a smaller slope value $\alpha$. The most realistic transient time can be attained by observing the slope value of the time evolution of the Hamming distance until it reaches a fixed value. Eventually, we find the values of the parameter $\alpha$ equal to $0.83\pm0.03$ and $0.91\pm0.03$ by implementing the damage spreading technique on the five different lattice sizes, $50\times50$, $75\times75$, $100\times100$, $200\times200$ and $400\times400$, of the isotropic and anisotropic BS models. The numerically obtained $\alpha$ values can be considered as a good estimation for $\alpha$ in $N\rightarrow\infty$ limit and together with the dynamical exponent $z$ lead to a data collapse of the finite size scaling for the five different lattice sizes of the isotropic and anisotropic BS models.

\begin{acknowledgement}
The authors wish to thank L. F. Lemmens for discussions and suggestions. This work has been supported by TUBITAK (Turkish Agency) under the Research Project number 104T148.
\end{acknowledgement}

%
\bibliographystyle{epj}
\bibliography{2dBS}
\end{document}